\documentclass{article}
\usepackage{arxiv}
\usepackage[utf8]{inputenc} % allow utf-8 input
\usepackage[T1]{fontenc}    % use 8-bit T1 fonts
\usepackage{url}            % simple URL typesetting
\usepackage{booktabs}       % professional-quality tables
\usepackage{amsfonts}       % blackboard math symbols
\usepackage{nicefrac}       % compact symbols for 1/2, etc.
\usepackage{microtype}      % microtypography
\usepackage{lipsum}
\usepackage{verbatim}
\usepackage{graphicx}
\usepackage{subfig}
\usepackage{xcolor}
\usepackage{floatrow}
\usepackage{tabstackengine}
\stackMath
\usepackage{setspace}
\usepackage{natbib}
\setcitestyle{authoryear}
\usepackage{amsthm}
\usepackage{amsmath}
\usepackage[colorlinks,citecolor=blue,urlcolor=blue,filecolor=blue,backref=page]{hyperref}
\usepackage{graphicx}
\usepackage{cancel}
\usepackage{float}
\usepackage{setspace}
\usepackage{epstopdf}
\usepackage{geometry}
\usepackage{graphics}
\restylefloat{table}
\usepackage{booktabs}%\linespread{1}
\usepackage[font=small,labelfont=bf,tableposition=top]{caption}
\DeclareCaptionLabelFormat{andtable}{#1~#2  \&  \tablename~\thetable}
\DeclareCaptionLabelFormat{andfigure}{#1~#2  \&  \figurename~\thefigure}
\newfloatcommand{capbtabbox}{table}[][\FBwidth]
\newcommand{\given}{\,|\,}

\title{A Compartment Model of Human Mobility and Early Covid-19 Dynamics in NYC}

\author{
    Ian Frankenburg \\
    UCLA Department of Biostatistics\\
    University of California, Los Angeles\\
    Los Angeles, CA 90095-1772 \\
    \texttt{ian.frankenburg@ucla.edu}
  \AND
  Sudipto Banerjee \\
  UCLA Department of Biostatistics\\
  University of California, Los Angeles\\
  Los Angeles, CA 90095-1772 \\
  \texttt{sudipto@ucla.edu}
}

\begin{document}
\date{January 26th, 2021}
\maketitle

\begin{abstract}
In this paper, we build a mechanistic system to understand the relation between a reduction in human mobility and Covid-19 spread dynamics within New York City. To this end, we propose a multivariate compartmental model that jointly models smartphone mobility data and case counts during the first 90 days of the epidemic. Parameter calibration is achieved through the formulation of a general Bayesian hierarchical model to provide uncertainty quantification of resulting estimates. The open-source probabilistic programming language Stan is used for the requisite computation. Through sensitivity analysis and out-of-sample forecasting, we find our simple and interpretable model provides evidence that reductions in human mobility altered case dynamics.
\end{abstract}


\section{Introduction}\label{Introduction}
\begin{figure}[H]
    \centering
    \subfloat{
        \includegraphics[width=70mm]{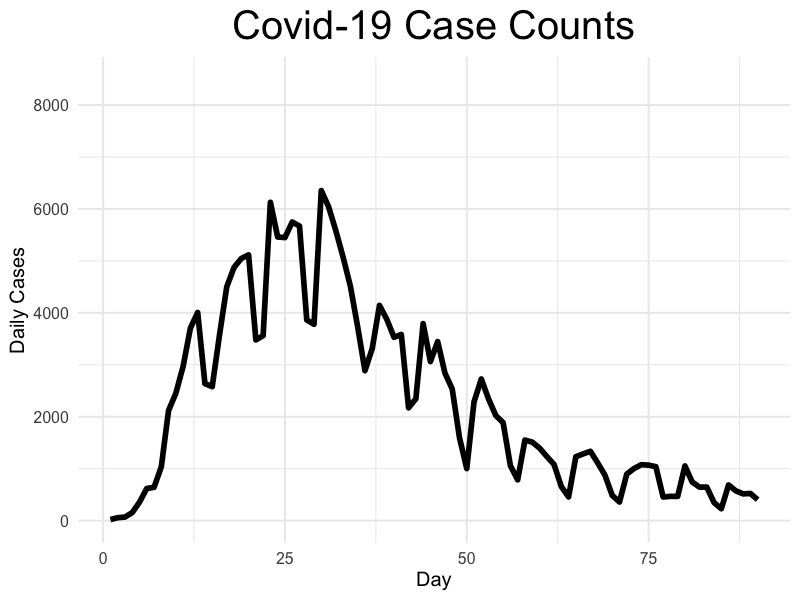}
    }
    \subfloat{
        \includegraphics[width=70mm]{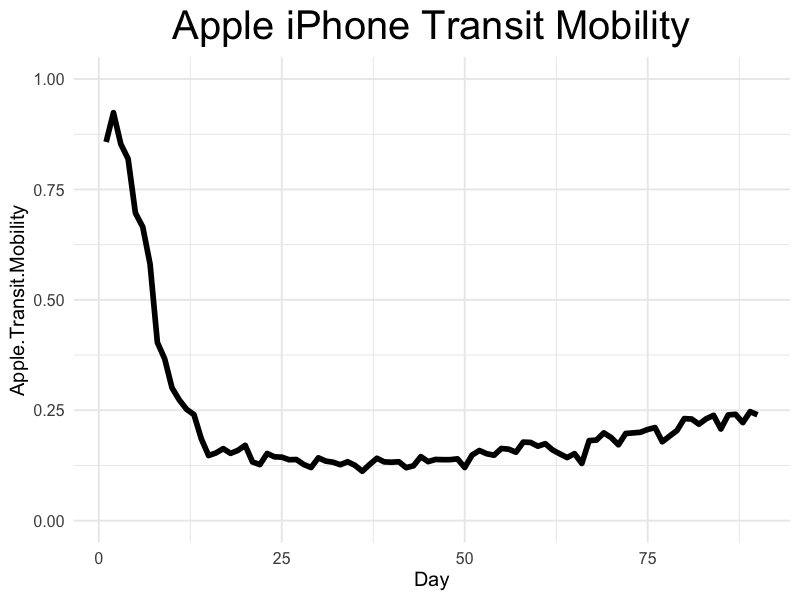}
    }
    \caption{Raw Daily Case Counts and Mobility Time Series}
    \label{fig:ts}
\end{figure}

\noindent 
The global Covid-19 pandemic has underscored the importance of mathematical and statistical models in understanding disease dynamics, assessing policy efficacy, and examining counterfactual scenarios to formulate thorough cost-benefit analyses. Lockdown measures can have drastic impact on individual well-being as well as society and the economy at large \citep{Bonaccorsi2020}, \citep{Cutler2020}. Therefore a retrospective study of Covid-19 lockdown and mitigation measures can help policymakers and public health officials understand to what end such efforts were effective. The formulation of a mechanistic compartmental model is a pathway towards such goals. In this article, we review compartmental model methodology, construct our new Bayesian hierarchical model, and discuss numerical methods relevant for implementation and fitting to real-world data. The new compartmental model is a simple modification of the classical susceptible-infectious-removed (SIR) model and enables a mechanistic correspondence between smartphone mobility data and infection dynamics. This can provide evidence of how reduced mobility due to early lockdowns or mitigation measures within New York City influenced Covid-19 spread dynamics. Case count data is obtained from the official website of the City of New York, available at \url{https://www1.nyc.gov/site/doh/covid/covid-19-data.page}. Population transit mobility data is obtained from \url{https://covid19.apple.com/mobility} and consists of anonymized Apple iPhone transit usage reported as a percent relative to baseline. Starting from the day after Governor Andrew Cuomo declared a state of emergency in New York State on March 7th, 2020, both the raw case count and transit mobility time series are presented below. Our end goal is then to establish a relationship between the two series shown in Figure \ref{fig:ts}.

\section{Background}
\label{background}
Mathematical modeling in epidemiology has a long history, famously dating back to the eighteenth century with the work of Bernoulli \citep{Dietz2002} or the mid-nineteenth century through John Snow's modeling of the cholera outbreak in London \citep{Shiode2015}. However, at the turn of the early twentieth century, mathematical epidemiology turned to the modern theory of dynamical systems analysis to understand outbreak evolution. In this section, we review the popular SIR model and subsequent mathematical analysis used to glean both qualitative and quantitative understanding of the dynamical system. This simple framework provides the necessary foundation for more complicated compartmental models with more population states. For examples of other compartment models designed to study early Covid-19 outbreak dynamics, see \citep{Hao2020} or \citep{Chaolong2020}.

\subsection{SIR Compartmental Model}
In modeling population-level data with a compartmental system, the population is typically subdivided into separate homogeneous groups. Here we focus on reviewing the simple SIR model developed in 1927 by  A. G. McKendrick and W. O. Kermack to model a plague outbreak in Bombay \citep{kermackProcRoySocLonA1927}. In such a model, the population is divided into susceptible $(S)$, infectious $(I)$, and removed $(R)$ groups. Individuals then progress through the various states at certain rates over time. The mathematical description of this changing system is the coupled set of ordinary differential equations
\begin{equation}
\label{sir}
    \begin{cases}
        \dfrac{dS(t)}{dt} &= -\beta S(t)I(t)/N\\[6pt]
        \dfrac{dI(t)}{dt} &= \beta S(t)I(t)/N -\gamma I(t)\\[6pt]
        \dfrac{dR(t)}{dt} &= \gamma I(t).
    \end{cases}
\end{equation}

The progression of the disease throughout the population depends upon the contact rate between susceptible and infectious individuals, the probability of transmission upon contact, and the prevalence of disease. To mathematically capture these factors and express the rate of new infections, let $\lambda$ be defined as a per capita contact rate among individuals. In this way, $\lambda S(t)$ will give the average number of susceptible contacts over time. Now let $p$ be the probability that a contact results in a new infection. Finally, the prevalence of the disease at time $t$ is by definition $I(t)/N$. Combining these terms gives $p\lambda S(t) I(t)/N$ as the incidence rate. In defining the \textit{effective contact rate} $\beta$ as the product of the per capita contact rate $\lambda$ and transmission probability $p$, the necessary form in equation (\ref{sir}) is recovered.

The remaining parameter $\gamma$ is interpreted by considering that $1/\gamma$ is the average sojourn time of an individuals within compartment $I$. It should also be noted the population is fixed throughout time since $N=S+I+R$. This can alternatively be seen since adding the terms in (\ref{sir}) gives zero. With meaning associated to the compartmental parameters, we next turn to an overview of the mathematical analysis involved in analyzing the basic SIR dynamical system.

\subsection{Linear Stability Analysis and the Basic Reproductive Number}
The system of differential equations in (\ref{sir}) are nonlinear, arising from the term $S(t)I(t)$. Linear stability analysis is the workhorse to understand the behavior of nonlinear dynamical systems and has a fundamental connection to the \textit{basic reproductive number} $\mathcal{R}_0$, popularized recently through media coverage of the Covid-19 pandemic. $\mathcal{R}_0$ is defined roughly to be the expected number of subsequent infections resulting from a single infected individual. In this section, we briefly review linear stability analysis and make the connection to $\mathcal{R}_0$. In the next section, we highlight the computation of $\mathcal{R}_0$ for a general class of compartmental models.

A steady state or equilibrium of a dynamical system is a point $\mathbf{x}^*$ where the system of differential equations evaluated at $\mathbf{x}^*$ is zero for all $t$. In this way, compartmental contents within the system are not changing over time. In the SIR model of (\ref{sir}), an important steady state is the so-called disease-free equilibrium of $\{(S^*,0,0): S^*\geq 0\}$. A natural subsequent question is the behavior of the system around small perturbations of the equilibrium. In computing the Jacobian about such a point, we linearize and are afforded tractable analysis. A heuristic justification arises from considering a Taylor expansion of the system about the disease-free equilibrium and ignoring high-order terms since the perturbation is assumed small. After dropping the explicit dependence on time $t$ from equation (\ref{sir}) to avoid clutter, the Jacobian of the system is computed as
\begin{equation}
    \mathbf J=\begin{pmatrix}
        \dfrac{\partial \dot S}{\partial S} & \dfrac{\partial \dot S}{\partial I} & \dfrac{\partial \dot S}{\partial R}\\[6pt]
        \dfrac{\partial \dot I}{\partial S} & \dfrac{\partial \dot I}{\partial I} & \dfrac{\partial \dot I}{\partial R}\\[6pt]
        \dfrac{\partial \dot R}{\partial S} & \dfrac{\partial \dot R}{\partial I} & \dfrac{\partial \dot R}{\partial R}\\[6pt]
    \end{pmatrix}=
    \begin{pmatrix}
        -\beta I/N & -\beta S/N & 0\\
        \beta I/N & \beta S/N-\gamma & 0\\
        0 & \gamma & 0\\[6pt]
    \end{pmatrix}.
\end{equation}
Through elementary matrix operations, $\mathbf J$ can be transformed to a triangular matrix. The eigenvalues are then found from inspection. Evaluating the Jacobian $\mathbf J$ at $(S^*,0,0)$ to linearize about the steady state results in eigenvalues of $\lambda_1=0$ and $\lambda_2=\beta S^*/N-\gamma$. The sign of these eigenvalues then determine the stability of the equilibrium point. The eigenvalue of $\lambda_1$ is ignored, as it corresponds to a line of equilibrium values $S^*$. In this way, the second eigenvalue of $\lambda_2$ is of main interest. If ${N}/{S^*}<{\beta}/{\gamma}$, then $\lambda_2>0$ and the steady state is unstable; otherwise it is stable. In epidemic terms, an outbreak occurs if the disease-free equilibrium is unstable. This ratio ${\beta}/{\gamma}$ acts as a bifurcation parameter in determining if an outbreak will occur and is thus afforded the fancy title of basic reproductive number. Letting the equilibrium point $S^*$ be the population size so that $S^*=N$, a simple relation emerges: if $\mathcal{R}_0>1$ the disease continues to spread but dies out otherwise. The effective reproductive number $\mathcal R_t$ then extends $\mathcal{R}_0$ by accounting for a changing susceptible population over time and is defined as $\mathcal R_t:=\mathcal{R}_0 S(t)/N$. A general method to compute $\mathcal R_0$ for more elaborate compartmental models will be discussed in the next subsection.

\subsection{Spectral Radius or Next Generation Matrix Method}
For general compartment models that extend the simplistic SIR framework, computing the basic reproductive number can be difficult. \citep{Diekmann2009} and \citep{Heffernan2005} describe a general method to compute $\mathcal{R}_0$ called the \textit{Next Generation Matrix} or \textit{Spectral Radius Method}, which we briefly review and compute for the SIR model.

Let $d\mathbf{X}(t)/dt$ represent a general coupled system of differential equations describing a compartmental model with $n$ components and $m$ infectious states. Define a vector-valued $\mathbf F(\mathbf{X}(t))$ to be a function where each component specifies flow rate into one of the respective $m$ infected compartments. Similarly, define a function $\mathbf{V}(\mathbf{X}(t))$ where each component specifies the flow rate out of a respective infectious compartment.

Next, $\mathbf F$ and $\mathbf V$ are linearized about the disease-free equilibrium point by computing the Jacobian. \citep{Diekmann2009} prove the Jacobian with respect to each infectious state will take the form
\begin{equation}
    \mathbf{J}(\mathbf{F})=\begin{pmatrix}
    \mathbf{A} & \boldsymbol{0}\\
    \boldsymbol{0} & \boldsymbol{0}
    \end{pmatrix}\text{ and }
    \mathbf{J}(\mathbf{V})=\begin{pmatrix}
    \mathbf{B} & \boldsymbol{0}\\
    \boldsymbol{*} & \boldsymbol{*}
    \end{pmatrix},
\end{equation}
where $\mathbf{A}$ and $\mathbf{B}$ are $m\times m$ matrices. The next generation matrix is then defined as $\mathbf{A}\mathbf{B}^{-1}$. The basic reproductive value of $\mathcal{R}_0$ is subsequently the spectral radius or largest eigenvalue of the next generation matrix. In the case of the SIR model, $\mathbf{F}(\mathbf{X}(t)):=-\beta S I/N$, while $\mathbf{V}(\mathbf{X}(t)):=\gamma I$. It follows that the necessary Jacobians evaluated at the disease-free equilibrium of $(N,0,0)$ results in $\mathbf{A}\mathbf{B}^{-1}={\beta}/{\gamma}$ and agree with the previous section.

\section{Methods}
\label{methods}
In this section, we detail the construction of our compartmental model designed to formulate an understanding of how reduction in human mobility might have altered early infection dynamics within New York City. The model is parsimonious, in that it consists of only four compartments and shares many well-established mathematical properties of the SIR model discussed in the background section. The dynamical system proposed comprises a closed population divided into susceptible, lockdown $(L)$, infectious, and removed states. Since the time horizon under investigation is short, we choose to ignore demographic factors such as birth, death, and migration. Below in Figure \ref{fig:flowDiag} is a visualization of the population progression through the different states.
\begin{figure}[H]
    \centering
    \includegraphics[width=70mm]{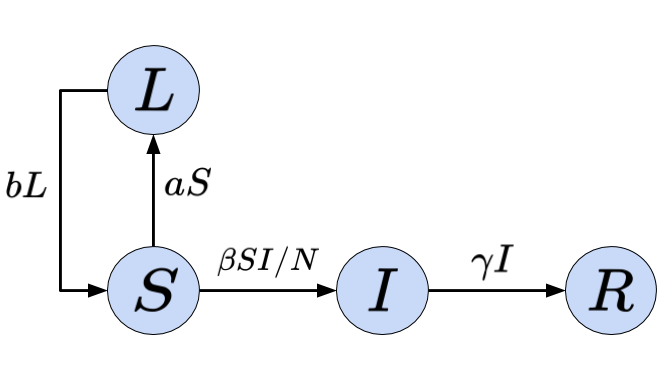}
    \caption{Proposed SLIR compartmental model}
    \label{fig:flowDiag}
\end{figure}
The system of differential equations describing the changing system are
\begin{equation}
    \label{eqn:diffEqs}
    \begin{cases}
        \dfrac{dS(t)}{dt} &= -\beta S(t)I(t)/N-aS(t)+bL(t)\\[6pt]
        \dfrac{dL(t)}{dt} &= aS(t)-bL(t)\\[6pt]
        \dfrac{dI(t)}{dt} &= \beta S(t)I(t)/N-\gamma I(t)\\[6pt]
        \dfrac{dR(t)}{dt} &= \gamma I(t),
    \end{cases}
\end{equation}
with initial conditions of $S(0)=N-i_0$, $L(0)=0$, $I(0)=i_0$, and $R(0)=0.$ An important qualitative feature of our model is that susceptible individuals are temporarily moved into the lockdown compartment to reflect social distancing, mitigation measures, and reduced mobility. Over time, individuals are reintroduced into the susceptible population out of the lockdown state. Through an application of the next generation matrix method described in section $\ref{background}$, $\mathcal{R}_0$ can be seen to be equivalent to the standard SIR model, i.e. $\mathcal{R}_0=\beta/\gamma$.

\subsection{A Bayesian Hierarchical Model}
We present our methodology in a general hierarchical framework to facilitate Bayesian inference of compartmental system parameters in equation (\ref{eqn:diffEqs}). In this hierarchical formulation, we can be explicit about the role of the mechanistic system, requisite numerical integration, and underlying process parameters. This section will construct the statistical model piece-by-piece. The hierarchy of connected components in the model can be visualised bottom-up as follows,
\begin{equation*}
 \text{System Parameters}\longrightarrow\text{Compartmental Mechanistic Model}\longrightarrow\text{Observed Data}.
\end{equation*}
We first establish notation to represent the mechanistic system of differential equations in the middle of the hierarchy. Let the equations in (\ref{eqn:diffEqs}) be denoted by $\mathbf{F}$, where
\begin{equation}\label{Xsolution}
\frac{d}{dt}\mathbf{X}(t) = \mathbf{F}(\mathbf{X}(t), t), \mbox{ where } \mathbf{X}(t) = \begin{pmatrix}S(t) \\ L(t) \\ I(t) \\ R(t) \end{pmatrix}\; \mbox{ and }\; 
    \mathbf{F}(\mathbf{X}(t), t) = 
    \begin{pmatrix}
    -\gamma\mathcal{R}_0 S(t)I(t)/N-aS(t)+bL(t)\\[4pt]
    aS(t)-bL(t)\\[4pt]
    \gamma\mathcal{R}_0 S(t)I(t)/N-\gamma I(t)\\[4pt]
    \gamma I(t)
    \end{pmatrix}
\end{equation}
and we have suppressed the dependence of $\{\mathcal{R}_0,\gamma,a,b\}$ in $\mathbf{F}(\mathbf{X}(t))$. The system is reparameterized in terms of $\mathcal{R}_0$ rather than $\beta$ to be more epidemiologically interpretable and results from a simple transformation of $\beta=\gamma\mathcal{R}_0$. To recover the system states $\mathbf X(t)$, the system of differential equations must be solved. Given fixed values of $\mathcal{R}_0$, $\gamma$, $a$, and $b$, the solution to the system of differential equations is a vector-valued function 
\begin{equation}
    \mathbf{X}(t)=\int\mathbf{F}(\mathbf{X}(t), t; \mathcal{R}_0,\gamma,a,b)dt.
\end{equation}
This solution will be necessary to connect with the observed data.

The top of the hierarchy is described by formulating a measurement process for the two outcome variables. Let the first outcome of interest be labeled ${Y}_L(t)$ and represent the percent of the population removed from the susceptible compartment by adhering to mitigation protocol.  The subscript $L$ is used for a reminder that this data is used to gain information on the lockdown compartment. Likewise, the second outcome is labeled ${Y}_I(t)$ and denotes the observed case counts over time. To model observation error in $Y_L(t)$, we choose a Beta distribution dependent upon a parameter $\phi_1$ to control dispersion, i.e.,
\begin{equation}
    Y_L(t)|L(t),\phi_1\sim\text{Beta}\big(\phi_1L(t)/N,\phi_1(1-L(t)/N)\big).
\end{equation}
Notice $L(t)$ is necessarily scaled by the population size $N$ to respect the support of the beta distribution. As we seek to inform the $L$ compartment through cell phone mobility data, the beta distribution is a natural choice because the data will be anonymized and reported as a percentage of nominal movement. This parameterization of the beta distribution has expectation and variance
\begin{align*}
    \mathbb{E}[Y_L(t)|L(t),\phi_1] &= L(t)/N\\
    \text{Var}(Y_L(t)|L(t),\phi_1) &= \frac{L(t)/N\big(1-L(t)/N\big)}{\phi_1+1}.
\end{align*}
To model observation noise in $Y_I(t)$, we use a negative binomial to account for overdispersion,
\begin{equation}
    Y_I(t)|I(t),\phi_1\sim\text{Negative Binomial}(I(t),\phi_2).
\end{equation}
Stan provides an alternative parameterization of the negative binomial called \texttt{neg\_binomial\_2}
with first two moments of
\begin{align*}
    \mathbb{E}[Y_I(t)|I(t),\phi_2] &= I(t)\\
    \text{Var}(Y_I(t)|I(t),\phi_2) &=  I(t) + \frac{I(t)^2}{\phi_2}.
\end{align*}
In this way, $\phi_2$ is viewed as a dispersion parameter. The full hierarchical description of the model can be completed by introducing prior distributions on the system parameters governing the differential equations. Writing the model in full,
\begin{equation}
    \label{hierarchy}
    \begin{split}
    {Y}_L(t)|L(t),\phi_1 &\sim \text{Beta}\big(\phi_1L(t)/N,\phi_1(1-L(t)/N)\big)\\
    {Y}_I(t)|I(t),\phi_2 &\sim \text{Negative Binomial}(I(t),\phi_2)\\
    \dfrac{d\mathbf{X}(t)}{dt} &= \mathbf{F}(\mathbf{X}(t), t; \mathcal{R}_0,\gamma,a,b)\\
    %{X}_S(t)|\mathcal{R}_0,\gamma,a,b &\sim \delta_{X_S}\big(\smallint\mathbf{F}(\mathbf{X}(t),t;\boldsymbol{\theta})dt\big)\\
    %{X}_I(t)|\mathcal{R}_0,\gamma,a,b &\sim \delta_{X_I}\big(\smallint\mathbf{F}(\mathbf{X}(t),t;\boldsymbol{\theta})dt\big)\\
    \mathcal{R}_0|\gamma&\sim \text{log-normal}(0,1)\\
    \gamma & \sim \text{Uniform}(0,1)\\
    \phi_1 & \sim \text{Inverse Gamma}(0.1,0.1)\\
    \phi_2 & \sim \text{Inverse Gamma}(0.1,0.1)\\
    a & \sim \text{Beta}(1,5)\\
    b & \sim \text{Uniform}(0,1).
    \end{split}
\end{equation}
The prior distributions on system parameters are weakly-informative. However, the prior distribution on $a$ might at first appear suspect. Through prior predictive checks, we find that placing a uniform prior on $a$ results in the SLIR model a priori favoring no epidemic breakout, as susceptibilities are removed from the population too quickly. Since the classical SIR model is a special case of our SLIR model as $a\to 0$, we place mass closer to 0 through the Beta(1,5) distribution to ensure the model generates reasonable predictions before seeing the data. The hierarchical model of the previous section crucially depends upon the numerical solution to a coupled set of differential equations of (\ref{eqn:diffEqs}). In the appendix section, we detail the internal workings of Stan's numerical optimization routines. Finally, efficient Bayesian analysis of parameters within the set of nonlinear differential equations relies upon the efficiencies gained through Hamiltonian Monte Carlo (HMC). A detailed review of this methodology is also included in the appendix section.

\section{Analysis and Results}
In this section, we present two simulation studies and conclude with the New York City analysis. First, the proposed SLIR compartmental model is used to simulate data from two lockdown scenarios that affect human mobility differently. We then fit our Bayesian model to assess whether the true parameter values are adequately recovered. After, we analyze the real-world mobility and case count data that initially inspired the model formulation.

\label{results}
\subsection{Simulated Data}
To illustrate the nonlinear dynamics of which our model can capture, the first simulation reflects the idealized scenario of strict adherence to lockdown and mitigation measures, when population movement is quickly reduced in the early stages of the outbreak and remains reduced for the next 90 days. In this case, individuals move from the $S$ compartment to the $L$ compartment quickly and are slowly reintroduced into the susceptible population so that population movement decreases to about 60\% relative to baseline within the first 20 days.

In the second scenario, we consider weak adherence to mitigation measures and illustrate the substantial change in dynamics by only altering the speed of flow back into the susceptible population from the $L$ compartment. In this case, the peak percentage of the population in the lockdown compartment is 30\% but quickly diminishes. In both simulations, the population size is fixed to $N=10,000$, $i_0=1$, and the true data-generating process is shown below as a dashed red line. The median of the posterior predictive distribution and 95\% credible intervals are shown in blue.

\begin{figure}[H]
\centering
\begin{tabular}{cc}
    \includegraphics[width=80mm]{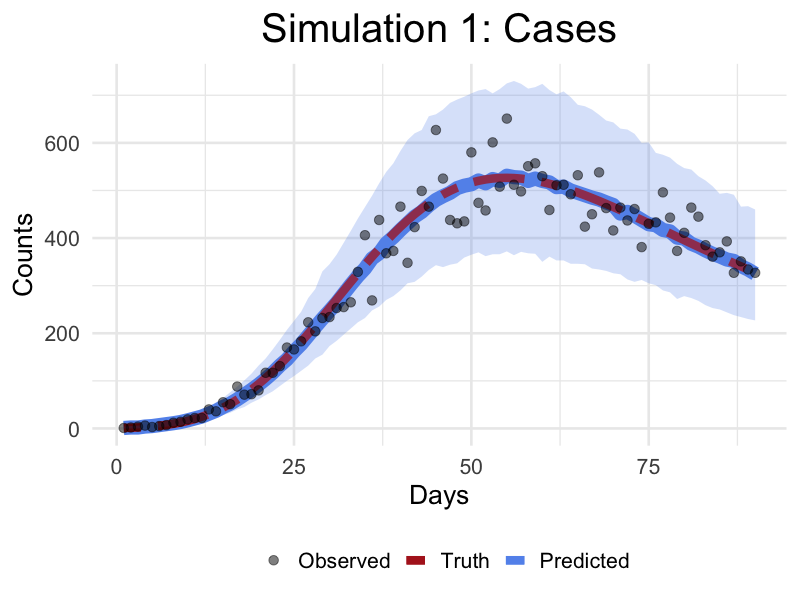} &   \includegraphics[width=80mm]{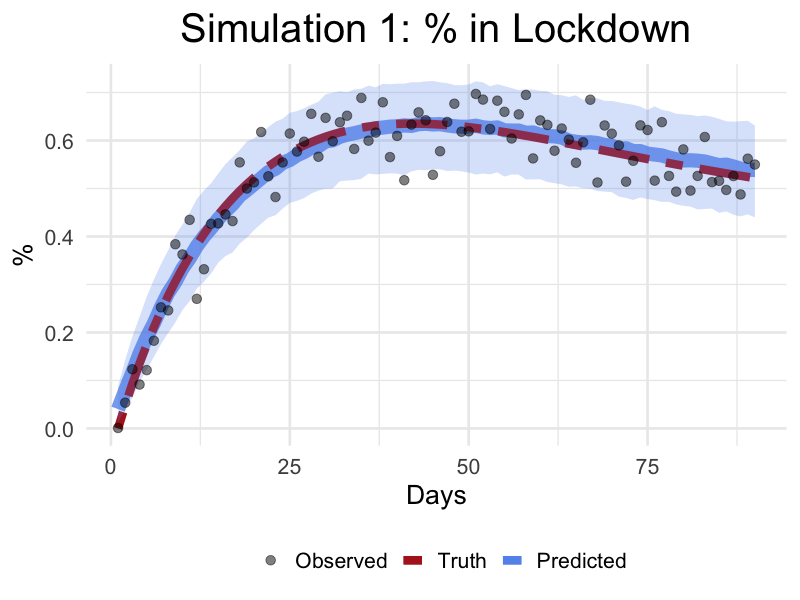} \\
    \includegraphics[width=80mm]{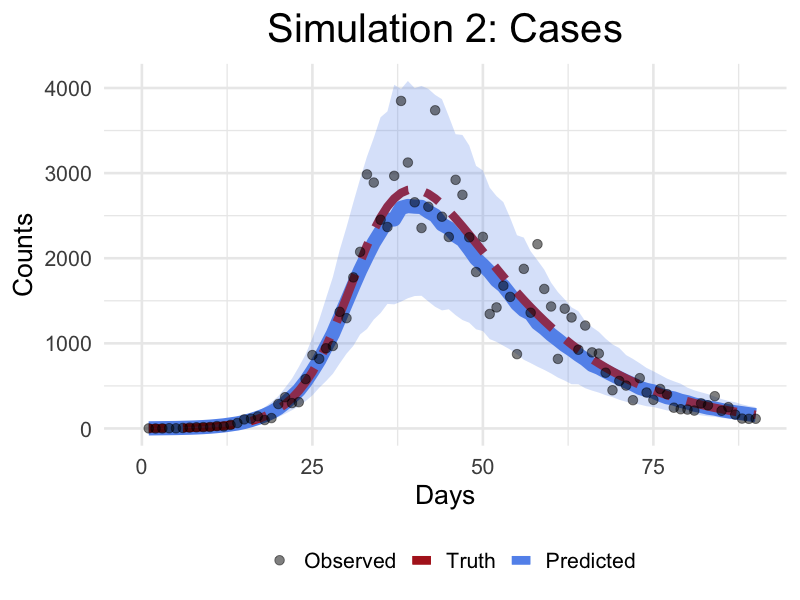} &   \includegraphics[width=80mm]{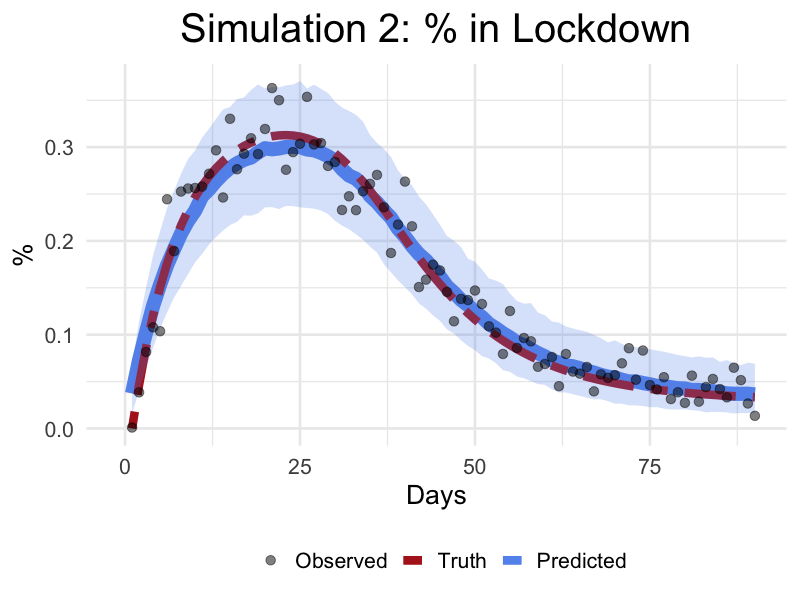}\\
    \end{tabular}
    \caption{Model Fit to Simulated Data}
\end{figure}

We fit our model to both scenarios using Stan's NUTS algorithm with 4 chains and 5,000 iterations each, the first half of which are discarded as warm-up.  The convergence of the parameter chains are judged by inspecting the trace plots along with the Gelman-Rubin $\hat R$ values, which compares the variation between chains to the variation within \citep{Gelman1992}. Ideally, the $\hat R$ value is close to one. These simulation results are displayed in in Table \ref{Rhat}.
\begin{table}[H] % t for top also possible <===========================
    \caption{Model Fit and Parameter Summaries} 
    \label{sequences}\centering
    \centering
    \begin{tabular}[t]{c c c c c c}
        \hline\hline \\ 
        Experiment & Parameter & Truth & Median & 95\% Credible Interval & G-R $\hat R$\\ [0.5ex]
        \hline \\ [-1.5ex]
        & $\mathcal{R}_0$ & 5 & 4.93 & 4.720 - 5.130 & 1.00\\ [0.5ex]
        Simulation 1 & $\gamma$ & 0.1 & 0.09 & 0.086 - 0.097 & 1.00\\ [0.5ex]
        & $a$ & 0.05 & 0.042 & 0.039 - 0.05 & 1.00\\ [0.5ex]
        & $b$ & 0.02 & 0.017 & 0.015 - 0.019 & 1.00\\ [0.2ex]
        \hline\\
        & $\mathcal{R}_0$ & 5 & 4.79 & 4.59 - 5.00 & 1.00\\ [0.5ex]
        Simulation 2 & $\gamma$ & 0.1 & 0.099 & 0.096 - 0.104 & 1.00 \\ [0.5ex]
        & $a$ & 0.05 & 0.045 & 0.04 - 0.051 & 1.00\\ [0.5ex]
        & $b$ & 0.1 & 0.095 & 0.086 - 0.106 & 1.00\\ [0.5ex]
        \hline
    \end{tabular}
    \label{Rhat}
\end{table}
In both cases, the Bayesian hierarchical model is able to infer the structural parameters of the SLIR model. Notice the change in case counts of both scenarios resulting from different susceptible population sizes.

\subsection{New York City Analysis}
The entire lead-up thus far was requisite background material for compartmental model inference and application to real-world data. As mentioned in the introduction, our main motivation for this article was to understand how a reduction in mobility affected early Covid-19 dynamics specifically within NYC. For convenience, we restate the data sources. Case counts are reported by the official website of the City of New York, available at \url{https://www1.nyc.gov/site/doh/covid/covid-19-data.page} and mobility data is hosted at \url{https://covid19.apple.com/mobility}. Since the Apple mobility data reflects a percent decrease in movement, it must first be transformed by subtraction from unity to adhere with the SLIR compartmental model structure. In other words, to prepare the mobility data for use in the hierarchical model, it must first be subtracted from one so that it no longer represents a percent decrease in transit mobility but rather a percent increase in individuals adhering to mitigation measures. Finally, we mention again that although the first case of Covid-19 in New York City was recorded on February 29th, we align our movement and case data to begin on March 8th, 2020, the day after Governor Andrew Cuomo declared a state of emergency in New York State. Finally, we take the initial number of cases to be the cases recorded on March 8th, 2020, and the population is fixed at 8,336,817 as determined by the US Census \citep{census:2019}.

To achieve parameter calibration, we fit in Stan our full hierarchical model using four chains run for 10,000 iterations each. We discard the first 5,000 as warm-up. Sufficient posterior exploration is assessed by examining parameter chain plots below and assessing $\hat R$ values, shown in Table \ref{table2}.

\begin{table}[H] % t for top also possible <===========================
    \caption{Model Fit and Parameter Summaries} 
    \label{table2}\centering
    \centering
    \begin{tabular}[t]{c c c c c c}
        \hline\hline \\ [-1.5ex]
        & Parameter & Median & 95\% Credible Interval & G-R $\hat R$\\ [0.5ex]
        \hline \\ [-1.5ex]
        & $\mathcal{R}_0$ & 5.13 & 4.841 - 5.448 & 1.00\\ [0.5ex]
        NYC Analysis & $\gamma$ & 0.212 & 0.191 - 0.234 & 1.00\\ [0.5ex]
        & $a$ & 0.115 & 0.106 - 0.124 & 1.00\\ [0.5ex]
        & $b$ & 0.0215 & 0.019 - 0.024 & 1.00\\
        \hline
    \end{tabular}
\end{table}

The fitted time series are presented below on the right, along with the raw data used to train the model. On the left, prior predictive distributions are included to illustrate the degree in which Bayesian learning occurs after observing the data. We also include the fit of an SIR model to illustrate its inability to capture the dynamics.
\begin{figure}[H]
    \centering
    \subfloat{
        \includegraphics[width=80mm]{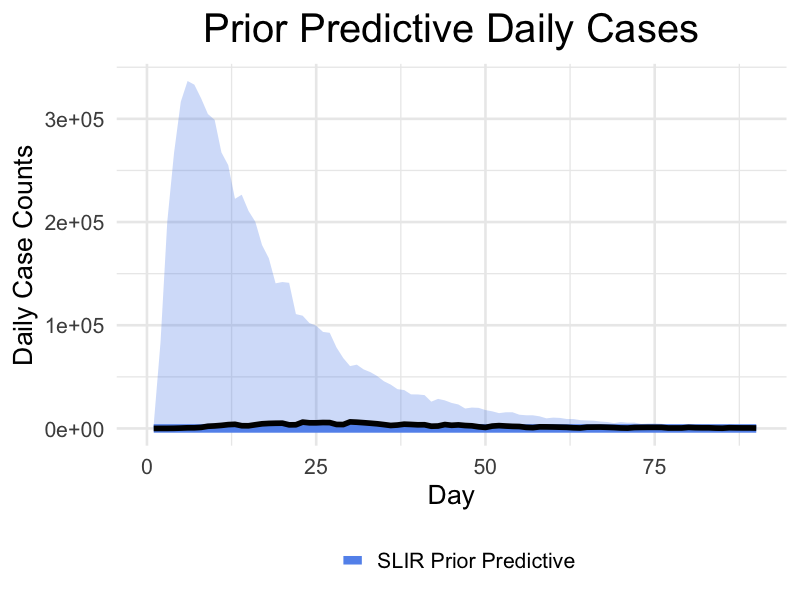}
    }
    \subfloat{
        \includegraphics[width=80mm]{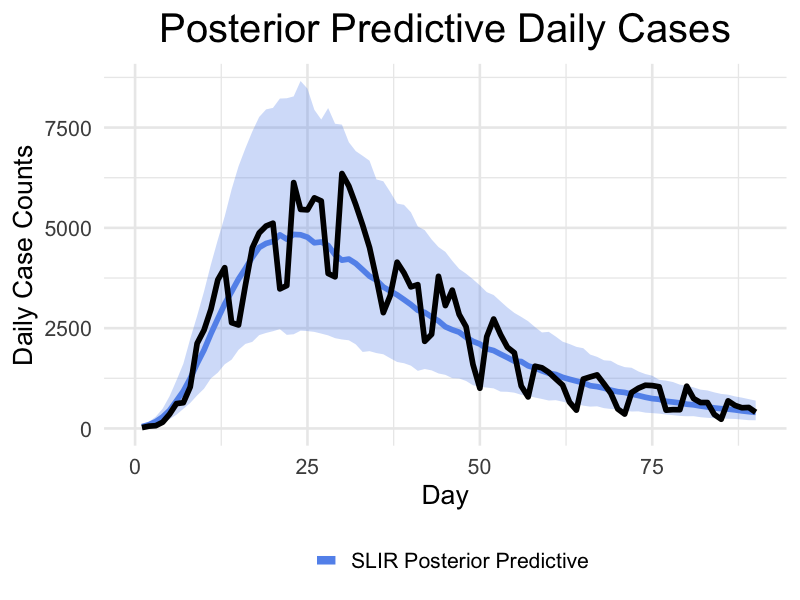}
    }
\end{figure}
\begin{figure}[H]
    \centering
    \subfloat{
        \includegraphics[width=80mm]{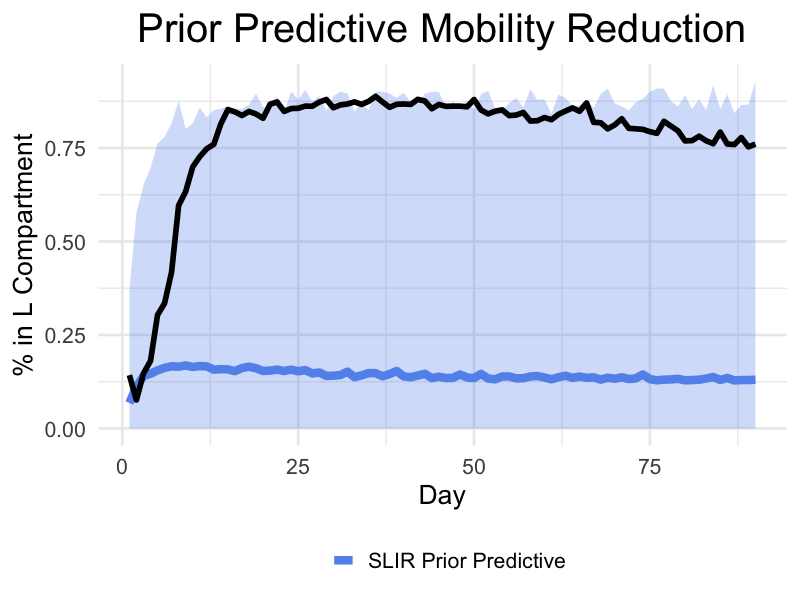}
    }
    \subfloat{
        \includegraphics[width=80mm]{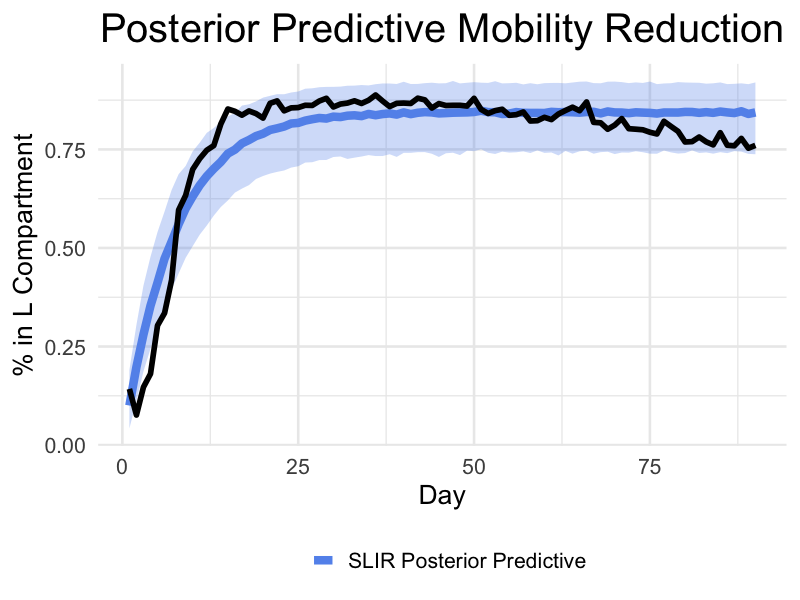}
    }
    \caption{SLIR Prior and Posterior Predictive Checks}
  \label{fig:fit}
\end{figure}

To assess the structural fit of our hypothesized SLIR mechanism, we next interpret parameter values to ensure they are logical and consistent with outside literature. The $\mathcal R_0$ estimate is cross-referenced with those of other popular online models. Using only death statistics as reported by Johns Hopkins University, \citep{covid_web} embeds a SEIR model in a machine learning framework for many regions across the United States and 70 countries. In this work, $\mathcal{R}_0$ is estimated for NYC to be between 5.0 and 5.8, in close agreement with our model. As an additional point of reference, \citep{Ives2020} provide an alternative methodology that fits a time series state-space model to death counts and explicitly accounts for reporting delays. $\mathcal{R}_0$ is estimated to be 6.3 with a 95\% confidence interval of 4.5-9. Our model thus has the added benefit of mechanistic interpretability as well as a tighter credible interval.

\begin{comment}
The SLIR estimate of $\mathcal{R}_0$ and the $\mathcal{R}_t$ curve are plotted below.
\begin{figure}[H]
    \centering
    \includegraphics[width=70mm]{images/r0_reff.png}
    \caption{$\mathcal{R}_0$ and $\mathcal{R}_t$}
\end{figure}
\end{comment}
The posterior of $\gamma$ has a median value of 0.21 and a 95\% credible interval of (0.191, 0.234). From this, we arrive at an estimate of approximately 5 days for the average infectious removal time. This estimate could be reasonable assuming asymptomatic or presymptomatic transmission is possible and that individuals isolate at the onset of symptoms; see \citep{Gandhi2020} for evidence. Outside work estimates symptom onset time to also be around 5 days. For example, \citep{Lauer2020} use 181 confirmed Covid-19 cases to estimate a median symptom onset time of 5.1 days with a 95\% confidence interval (4.5, 5.8) days. A systematic review by \citep{Madjid2020} estimates a mean symptom onset time of 5.2 days with a 95\% confidence interval of (4.1, 7.0) days. These estimates provide credence in establishing a correspondence between population movement reduction and infection dynamics with the mechanistic SLIR model.  

\subsection{Sensitivity Analysis and Out-of-sample Forecasting}
\begin{figure}[!ht]
    \centering
    \begin{tabular}[b]{cccc}
        \hline\hline \\ [-1.5ex]
        & \% Decline in Mobility & \% Pop. Infected\\ [1.5ex]
        \hline \\ [1.5ex]
        & 100\% & 2.4\% \\ [1.5ex]
        & \color{blue}{80}\% & \color{blue}{2.5}\% \\ [1.5ex]
        & 60\% & 18\% \\ [1.5ex]
        & 40\% & 84\% \\ [1.5ex]
        & 20\% & 95\% \\ [1.5ex]
        \hline\\
        \\
        \\
        \\
    \end{tabular}
    \qquad
    \includegraphics[scale=.3]{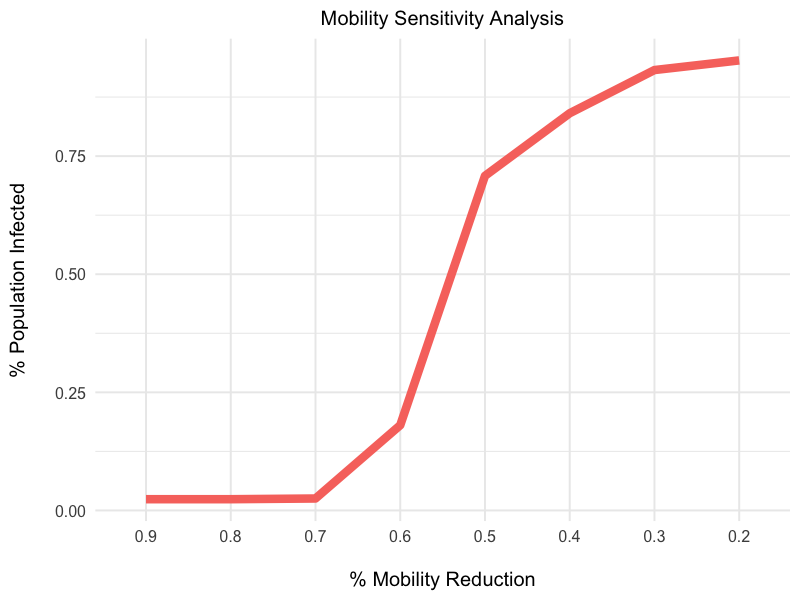}%
    \captionlistentry[table]{A table beside a figure}
    \captionsetup{labelformat=andtable}
    \caption{Mobility Reduction vs. \% Population Infected}
    \label{fig:curve}
\end{figure}

We conclude the New York City analysis by assessing the sensitivity of the model to changing mobility levels. Additionally, we assess out-of-sample predictive capacity of the SLIR model. To perform the sensitivity analysis, we first generate a range of hypothetical mobility scenarios, from full mobility reduction through extremely stringent lockdown measures to a more mild decline. This is displayed in Figure \ref{fig:curve}, with the true, real-world observed mobility levels highlighted as blue. 

\begin{comment}
\begin{figure}[H]
    \begin{floatrow}
        \capbtabbox{%
            \begin{tabular}[b]{cccc}\hline
                \hline\hline \\ [-1.5ex]
                & \% Decline in Mobility & \% Pop. Infected\\ [0.5ex]
                \hline \\ [-1.5ex]
                & 100\% & 2.4\% \\
                & \color{blue}{80}\% & \color{blue}{2.5}\% \\ [0.5ex]
                & 60\% & 18\% \\ [0.5ex]
                & 40\% & 84\% \\ [0.5ex]
                & 20\% & 95\% \\ [0.5ex]
                \hline
            \end{tabular}
        }{%
          \caption{Estimated \% Population Infected}%
        }\newline
        \ffigbox{%
          \includegraphics[scale=.3]{arxiv/images/sens.png}%
        }{%
          \caption{Infection/Mobility Curve}%
          \label{fig:curve}
        }
    \end{floatrow}
\end{figure}
\end{comment}

It is important to note the explosive case growth with mobility reduction levels under 60\% due to the nonlinear dynamics present in SIR-type models. This is evidence that a mobility reduction significantly altered infections within the city, assuming SIR-type dynamics.

\begin{figure}[H]
    \centering
    \includegraphics[scale=.35]{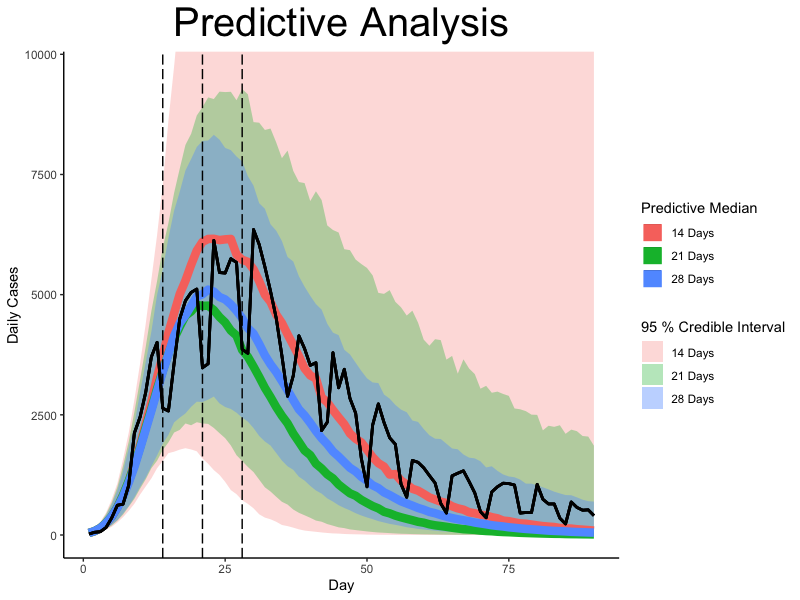}
    \caption{SLIR Out-of-sample Forecasts}
    \label{fig:forecast}
\end{figure}

To conclude this section, we analyse the out-of-sample forecasting ability of the SLIR model when trained on only a subset of the ninety day period. We consider for illustration a two, three, and four week training intervals. These are indicated by vertical dashed lines in Figure \ref{fig:forecast}.
The predictive median is illustrated with a solid line. With only two weeks observed, the predictive median is reasonably representative of the future trajectory but with very large uncertainty. The upper 95\% predictive curve for the two week window reaches roughly 400,000 infections, but is clear that the predictive interval quickly contracts as more data is observed (cf. Figure \ref{fig:fit}). Both the three week and four week training periods have contracted credible interval forecasts that contain the observed data. Finally, we mention that we were unable to fit the standard SIR model to the New York City data. Using the quasi-Newton optimization functions available in Stan, we found the SIR model fit was highly unstable across a range of starting values indicating extreme multi-modality in the likelihood surface. 

\section{Discussion and Future Work}

\subsection{New York City Analysis}
In this work, we have formulated a basic extension of the classical SIR model to jointly fit cell phone transit mobility data and case counts. By jointly modeling two outcome variables, we establish a mechanistic correspondence between reduced mobility and infection dynamics. The applied analysis and findings, however, are limited to NYC during the first 90 days. The disease progression throughout the region was well-approximated by deterministic dynamics and modeling with a simple compartment system. In other geographic regions, the dynamics might not be as suitably well-behaved or understood. It is difficult to capture the myriad of factors contributing to disease spread throughout a population, and often a stochastic model may be more appropriate. Additionally, the time horizon considered in our applied analysis is relatively short. As a disease progresses throughout the population and becomes more widespread, the strategy of informing the susceptible population numbers through cell phone mobility data becomes more difficult. It is also worth consideration about whether in general smartphone mobility data can be considered a representative sample of individuals' adherence to mitigation protocols. It is therefore difficult to build a quantitative understanding of the dynamics between lockdown measures and disease progression over long periods of time. A direction for future research is to establish modeling strategies for such scenarios. This will introduce additional challenges of accounting for population immunity, vaccinations, and natural seasonality effects.

\subsection{Stochastic Model Extensions}
We briefly discuss for both discrete and continuous time how the deterministic SLIR model can be relaxed through a state-space (or Hidden Markov Model) framework, where stochasticity is introduced into the underlying driving epidemic process. We first discuss a discrete-time extension by modifying equation (\ref{hierarchy}) so that the numerical solution to $\frac{d\mathbf{X}(t)}{dt}$ is embedded within a likelihood function at the discrete observation time points. In this way, equation (\ref{hierarchy}) is modified to include $p(\mathbf{X}(t)|\boldsymbol\theta,\mathcal{R}_0,\gamma,a,b)$, where $\boldsymbol\theta$ is introduce to accommodate new parameters. In other words, stochasticity is introduced to the driving dynamics through the choice of suitable probability distribution that is a function of the numerical solution of the system of differential equations. This approach has been successfully applied to forecast seasonal influenza \citep{statespace2017}, \citep{statespace2012}, as well as to assess Covid-19 interventions in China \citep{Wang2020} and Japan \citep{Kobayashi2020}. 

An alternative approach is to incorporate continuous-time stochasticity directly by expressing $d\mathbf{X}(t)$ as a stochastic differential equation (SDEs) \citep{oksendal2003stochastic}. In this case, one modification of equation (\ref{hierarchy}) could be
\begin{equation}
    \label{sde}
    d\mathbf{X}(t) = \boldsymbol\mu(\mathbf{X}(t), t)dt + \mathbf L(\mathbf{X}(t), t)d\mathbf W(t),
\end{equation}
where $\mathbf{W}(t) = (W_1(t),W_2(t),W_3(t),W_4(t))^\top$ is a vector of standard Wiener processes or Brownian motions, $\boldsymbol{\mu}$ is some vector-valued function, and $\mathbf{L}$ is a compatible matrix. Depending on the structure of $\boldsymbol{\mu}$, there may or may not exist an explicit solution. A SDE that affords a closed form solution is the Ornstein-Uhlenbeck (OU) process. See \citep{Chatzilena2019} for Stan code in the case where the infectious compartment of the SIR model is endowed an OU process to allow for random fluctuations. See also \citep{Wang2018} for an analysis of the susceptible-infectious-susceptible model where time-varying parameters are introduced through an OU process. In both examples, the analytic nature of the solution to the SDE enables efficient implementation in Stan or alternative software. In cases where no closed-form solution to (\ref{sde}) exists, inference in such a system is closely related to Bayesian filtering and smoothing, amenable to such methods as the Kalman filter and extensions \citep{sarkka_solin_2019}. The New York City data was captured well by deterministic dynamics, but often elsewhere disease progression is not suitably captured within such a framework. Bayesian inference for SDEs modeling infectious disease dynamics is thus an attractive area of future research.

\section{Appendix}
\subsection{Numerical Solvers}
The hierarchical model of the previous section crucially depends upon the numerical solution to a coupled set of differential equations. We briefly review the popular numerical integration schemes and connect them with our hierarchical model of the compartmental system in (\ref{eqn:diffEqs}). In practice, solutions to the SIR model are numerically approximated and entail specific computational challenges. Runge-Kutta (RK) is a classical numerical method \citep{Struthers2019, numericsBook} in which we employ to numerically solve this set of nonlinear differential equations. RK generalizes the well-known Euler method for iteratively solving systems of differential equations. In the following, we formulate the RK method in vector notation for the proposed SLIR compartmental model.

Consider $\mathbf{X}(t)$ in (\ref{Xsolution}). From the fundamental theorem of Calculus, we know that
\begin{equation}\label{eqn:FunThmCalc}
        \mathbf{X}(t+\Delta t)=\mathbf{X}(t)+\int_{t}^{t+\Delta t}\mathbf{F}(\mathbf{X}(u), u)du\;.
\end{equation}
Given the initial value $\mathbf{X}(t_0)$ at time $t_0$, numerically solving for $\mathbf{X}(t)$ amounts to constructing a sequence $\{\mathbf{X}_n : n=0,1,\ldots,T\}$, where $\mathbf{X}_n := \mathbf{X}(t_n)$ and $t_n = t_{n-1} + \Delta t$ are equispaced time points for $n=0,1,\ldots,T$, by approximating the integral of $\mathbf{F}(\mathbf{X}(t),t)$ in (\ref{eqn:FunThmCalc}). Using this approximation, a sequence is generated starting from some $t_0$ as,
\begin{equation}\label{sequence}
    \begin{split}
    \mathbf{X}_{n+1} = \mathbf{X}_n + \int_{t_n}^{t_n+\Delta t}\mathbf{F}(\mathbf{X}(u), u)du\;.    
    \end{split}
\end{equation}
where the initial condition $\mathbf{X}_0:=\mathbf{X}(t_0)$ is given. There are several customary choices for such approximations, but for most practical purposes one need not look beyond the following:
\begin{equation}\label{eqn:quadrature}
 \int_{t}^{t+\Delta t}\mathbf{F}(\mathbf{X}(t),t)dt =
 \left\{ 
 \begin{array}{cl}
  (\Delta t) \mathbf{F}(\mathbf{X}(t),t) & \mbox{(Euler)}\\
  \frac{(\Delta t)}{2}\left(\mathbf{F}(\mathbf{X}(t),t) + \mathbf{F}(\mathbf{X}(t+\Delta t),t+\Delta t)\right) & \mbox{(Trapezoidal)} \\
  (\Delta t) \mathbf{F}(\mathbf{X}(t + (\Delta t)/2), t + (\Delta t)/2) & \mbox{(Midpoint or Modified Euler)}\;.
 \end{array}
 \right.
%  \begin{cases}
%  (\Delta t) \mathbf{F}(\mathbf{X}(t),t)\quad \mbox{(Euler)} \\
%  (\Delta t) \frac{\mathbf{F}(\mathbf{X}(t),t) + \mathbf{F}(\mathbf{X}(t+\Delta t),t+\Delta t)}{2}\quad \mbox{(Trapezoidal)} \\
%  (\Delta t) \mathbf{F}(\mathbf{X}(t + (\Delta t)/2), t + (\Delta t)/2)\quad \mbox{(Midpoint or Modified Euler)}
%  \end{cases}
\end{equation}
Euler's approximation (first equation in (\ref{eqn:quadrature})) is the simplest of the approximations which yields Euler's Method. In this case, the sequence is constructed as 
\begin{equation}\label{eqn:euler}
 \mathbf{X}_{n+1} = \mathbf{X}_n + (\Delta t) \mathbf{F}(\mathbf{X}_n, t_n)\; \mbox{ for }\; n=0,1,\ldots,T-1\;. 
\end{equation}
Starting with $\mathbf{X}_0$, each element of the sequence in (\ref{eqn:euler}) is computed since $\mathbf{F}(\mathbf{X}_n, t_n)$ is available. The spacing between the time points, $\Delta t$, is specified by the user and controls the resolution of the numerical solution.

Euler's approximation is easy to execute and based upon a first order Taylor expansion. It is also the least accurate. The Trapezoidal and Modified Euler approximations in (\ref{eqn:quadrature}) are based upon numerical integration using trapezoidal areas and the midpoint approximation. Both these methods help improve Euler's method with the Modified Euler outperforming the trapezoidal rule in terms of accuracy. 
\begin{comment}
by seeking to iterate  
\begin{equation}\label{eqn:heun-modified-euler}
 \mathbf{X}_{n+1} = \left\{\begin{array}{cl}
                             \mathbf{X}_n + (\Delta t)\frac{\mathbf{F}(\mathbf{X}_n,t_n) + \mathbf{F}(\mathbf{X}_{n+1},t_{n+1})}{2} & \mbox{(Trapezoidal)}\;; \\
                             \mathbf{X}_n + (\Delta t)\left(\mathbf{F}(\mathbf{X}_{n+1/2}, t_{n+1/2}\right) & \mbox{(Modified Euler)}
                            \end{array}\;\right.
  \mbox{ for }\; n=0,1,\ldots,T-1\;,
\end{equation}
where $\mathbf{X}_{n+1/2} := \mathbf{X}(t_n + (\Delta t)/2)$ and $t_{n+1/2} = t_n + (\Delta) t/2$. 
\end{comment}
However, the Trapezoidal and the Modified Euler methods involve $\mathbf{X}_{n+1}$ and $\mathbf{X}_{n+1/2} := \mathbf{X}(t_{n+1/2})$ with $t_{n+1/2} := t_n + (\Delta) t/2$, respectively, which are unknown at iteration $n$. In fact, $\mathbf{X}_{n+1}$ is the very quantity we wish to compute at iteration $n$. Therefore, we substitute these unknown quantities with their first order (Euler) approximations in (\ref{eqn:euler}) that are available at iteration $n$. For each $n$ we compute $\displaystyle \mathbf{a}_n = \mathbf{F}\left(\mathbf{X}_n, t_n\right)$, $\displaystyle \mathbf{b}_n = \mathbf{F}\left(\mathbf{X}_n + (\Delta t)\mathbf{a}_n, t_{n+1}\right)$ and $\displaystyle \mathbf{c}_n = \mathbf{F}\left(\mathbf{X}_n + (\Delta t / 2)\mathbf{a}_n, t_{n+1/2}\right)$ and transition from $\mathbf{X}_n$ to $\mathbf{X}_{n+1}$ as
\begin{equation}\label{eqn:heun-modified-euler-updates}
 \mathbf{X}_{n+1} = \left\{\begin{array}{cl}
                             \mathbf{X}_n + (\Delta t)\frac{\left(\mathbf{a}_n + \mathbf{b}_n\right)}{2} & \mbox{(Trapezoidal)}\;; \\
                             \mathbf{X}_n + (\Delta t)\mathbf{c}_n & \mbox{(Modified Euler)}
                            \end{array}\;\right.
  \mbox{ for }\; n=0,1,\ldots,T-1\;,
\end{equation}
Both methods in (\ref{eqn:heun-modified-euler-updates}) deliver noticeable improvements over Euler's method in (\ref{eqn:euler}). Numerical error from (\ref{eqn:heun-modified-euler-updates}) are much smaller in magnitude and grow less quickly. This can be explained by observing that while (\ref{eqn:euler}) depends only upon the data available at $t_n$ (only one data point), the two methods in (\ref{eqn:heun-modified-euler-updates}) use current data at $t_n$ along with estimates of the slope at a point that lies in the future. While these estimates are computed using only the currently available data, they still produce substantially improved estimates. Also see \cite{TREIBER2015183} for comparisons among different numerical integration schemes.

Higher order Taylor expansions produce other iterative schemes. Thus, second order methods emerge from
\begin{equation}\label{eqn:taylor}
 \mathbf{X}(t + \Delta t) = \mathbf{X}(t) + (\Delta t)\mathbf{F}(\mathbf{X}(t),t) + \frac{(\Delta t)^2}{2}\frac{d}{dt}\mathbf{F}(\mathbf{X}(t),t) + O((\Delta t)^2)\; .
\end{equation}
A second order iterative scheme corresponding to (\ref{eqn:taylor}) updates
\begin{equation}\label{eqn:taylor_updates}
 \mathbf{X}_{n+1} = \mathbf{X}_n + (\Delta t)\mathbf{F}(\mathbf{X}_n,t_n) + \frac{(\Delta t)^2}{2}\left\{{\left.\frac{\partial \mathbf{F}}{\partial t}\right|_{t=t_n}} + {\left. \left[\frac{\partial \mathbf{F}}{\partial \mathbf{X}}\right]\right|_{\mathbf{X}=\mathbf{X}_n}}{\left.\frac{d}{dt}\mathbf{X}(t)\right|_{t=t_n}} \right\} \;,
\end{equation}
where we have used the multivariable chain-rule of derivatives to evaluate the derivative of $\mathbf{F}(\mathbf{X}(t),t)$ with respect to $t$, $\displaystyle \left[\frac{\partial \mathbf{F}}{\partial \mathbf{X}}\right]$ is the matrix with $(i,j)$-th element being the partial derivative of the $i$-th element of $\mathbf{F}(\mathbf{X}(t),t)$ with respect to the $j$-th variable in $\mathbf{X}$, and $\displaystyle \left.\frac{d}{dt}\mathbf{X}(t)\right|_{t=t_n} = \mathbf{F}(\mathbf{X}_n, t_n)$. Unfortunately, computing (\ref{eqn:taylor_updates}) requires the derivatives of $\mathbf{F}(\mathbf{X}(t),t)$ and may, in general, be numerically cumbersome.

The Runge-Kutta methods are among the most conspicuous of numerical methods for solving systems of ordinary differential equations. The underlying idea is to achieve the same accuracy as Taylor series updates without requiring higher order derivatives of $\mathbf{F}(\mathbf{X}(t))$. We can motivate this approach from the earlier methods. In (\ref{eqn:heun-modified-euler-updates}), the Trapezoidal and Modified Euler methods define the updates using $\mathbf{a}_n$, $\mathbf{b}_n$ and $\mathbf{c}_n$ that are completely specified. In particular, observe that the Trapezoidal method updates using a weighted average of $\mathbf{a}_n$ and $\mathbf{b}_n$. Instead of prescribing $\mathbf{a}_n$ and $\mathbf{b}_n$, the Runge-Kutta approach prefers to find weighted averages to ensure that the approximation matches that from a Taylor series expansion such as in (\ref{eqn:taylor}) or (\ref{eqn:taylor_updates}). Therefore, a second-order Runge-Kutta method (RK2) writes
\begin{equation}\label{eqn:rk2_updates}
 \mathbf{X}_{n+1} = \mathbf{X}_n + (\Delta t)\left\{\omega_1\mathbf{a}_n + \omega_2\mathbf{F}(\mathbf{X}_n + (\Delta t)\beta\mathbf{a}_n, t_n + (\Delta t)\alpha)\right\} 
\end{equation}
and seeks to find $\omega_1$, $\omega_2$, $\alpha$ and $\beta$ so that the approximation matches (\ref{eqn:taylor_updates}). Substituting the first-order expansion,
\begin{equation}
 \begin{split}
  \mathbf{F}(\mathbf{X}_n &+ (\Delta t)\beta\mathbf{a}_n, t_n + \alpha (\Delta t)) = \mathbf{F}(\mathbf{X}_n, t_n) + (\Delta t)\beta{\left.\left[\frac{\partial \mathbf{F}}{\partial \mathbf{X}}\right]\right|_{\mathbf{X}=\mathbf{X}_n}}\mathbf{a}_n + (\Delta t)\alpha{\left.\frac{\partial \mathbf{F}}{\partial t}\right|_{t=t_n}} + O((\Delta t)^2) \\
  &= \mathbf{a}_n + (\Delta t)\beta{\left.\left[\frac{\partial \mathbf{F}}{\partial \mathbf{X}}\right]\right|_{\mathbf{X}=\mathbf{X}_n}}\mathbf{a}_n + (\Delta t)\alpha\left\{{\left.\frac{\partial \mathbf{F}}{\partial t}\right|_{t=t_n}} + \left[\frac{\partial \mathbf{F}}{\partial \mathbf{X}}\right]{\left.\frac{d}{dt}\mathbf{X}(t)\right|_{t=t_n}} \right\} + O((\Delta t)^2)
  \;, 
 \end{split}
\end{equation}
into the right hand side of (\ref{eqn:rk2_updates}) and comparing with the expansion (\ref{eqn:taylor_updates}) we find that the two expansions are equivalent if
\begin{equation}
\omega_1 + \omega_2 = 1\;;\quad \omega_2\beta = \omega_2\alpha = 1/2\;.  
\end{equation}
RK2 specifies $\omega_1=\omega_2=1/2$ and $\alpha=\beta=1$, which, when substituted into (\ref{eqn:rk2_updates}), yields the Trapezoidal approximation.

More generally, the explicit RK methods of order $s$ specify updating schemes
\begin{equation}\label{eqn:rk_general}
 \mathbf{X}_{n+1} = \mathbf{X}_n + (\Delta t)\sum_{i=1}^s\omega_i \mathbf{k}_i\;,
\end{equation}
where $\mathbf{k}_1 = \mathbf{F}(\mathbf{X}_n, t_n + (\Delta t)\alpha_1)$, and $\mathbf{k}_i = \mathbf{F}\left(\mathbf{X}_n + (\Delta t)\sum_{j=1}^{i-1}\beta_{ij}\mathbf{k}_j, t_n + (\Delta t)\alpha_i\right)$ for $i=2,\ldots,s$. The coefficients are found from an $s$-th order Taylor expansion. A popular choice sets $\alpha_1=0$ and solves
\begin{equation}\label{eqn:rk_general_coefficients}
 \sum_{i=1}^s \omega_i = 1\; \mbox{ and }\; \sum_{j=1}^{i-1}\beta_{ij} = \alpha_i\; \mbox{ for }\; i=2,3,\ldots,s\;.
\end{equation}
In particular, the RK4 method specifies the following values in (\ref{eqn:rk_general}):
\begin{equation}\label{eqn:rk4_coefficients}
 s=4\;;\quad \begin{bmatrix} \omega_1 \\ \omega_2 \\ \omega_3 \\ \omega_4 \end{bmatrix} = \frac{1}{6}\begin{bmatrix} 1 \\ 2 \\ 2 \\ 1\end{bmatrix}\;;\quad \begin{bmatrix} \beta_{21} & & \\ \beta_{31} & \beta_{32} & \\ \beta_{41} & \beta_{42} & \beta_{43} \end{bmatrix} = \begin{bmatrix} \frac{1}{2} & & \\ 0 & \frac{1}{2} & \\ 0 & 0 & 1 \end{bmatrix}\;;\quad \begin{bmatrix} \alpha_1 \\ \alpha_2 \\ \alpha_3 \\ \alpha_4 \end{bmatrix} = \frac{1}{2}\begin{bmatrix} 0 \\ 1 \\ 1 \\ 2 \end{bmatrix}\;.                                                                                                                                                                        
\end{equation}

\begin{comment}
\begin{equation}
\mathbf{X}_{n+1} = \mathbf{X}_n + \frac{1}{6}\left(\mathbf{k}_1 + 2\mathbf{k}_2 + 2\mathbf{k}_3 + \mathbf{k}_4\right)\;,
%     \begin{aligned}
%     \label{rk4}
%         \widehat{\mathbf{X}}_1(t)&=\mathbf{F}(\widehat{\mathbf{X}}_1(t), t)\Delta t\\
%         \widehat{\mathbf{X}}_2(t)&=\mathbf{F}(\widehat{\mathbf{X}}(t)+\widehat{\mathbf{X}}_1(t)/2, t+\Delta t/2)\Delta t\\
%         \widehat{\mathbf{X}}_3(t)&=\mathbf{F}(\widehat{\mathbf{X}}(t)+\widehat{\mathbf{X}}_2(t)/2, t+\Delta t/2)\Delta t\\
%         \widehat{\mathbf{X}}_4(t)&=\mathbf{F}(\widehat{\mathbf{X}}(t)+\widehat{\mathbf{X}}_3(t), t+\Delta t)\Delta t.
%     \end{aligned}
\end{equation}
where
\begin{equation}
 \begin{split}
  \mathbf{k}_1 &= \mathbf{F}\left(\mathbf{X}_n, t_n\right)\;; \\
  \mathbf{k}_2 &= \mathbf{F}\left(\mathbf{X}_n + (\Delta t)\frac{\mathbf{k}_1}{2}, t_n + \frac{(\Delta t)}{2}\right)\;; \\
  \mathbf{k}_3 &= \mathbf{F}\left(\mathbf{X}_n + (\Delta t)\frac{\mathbf{k}_2}{2}, t_n + \frac{(\Delta t)}{2}\right)\;; \\
  \mathbf{k}_4 &= \mathbf{F}\left(\mathbf{X}_n + (\Delta t)\mathbf{k}_3, t_n + (\Delta t)\right)\;; \\
 \end{split}
\end{equation}
\end{comment}
The appropriate step size $\Delta t$ is hard to determine. An advantage of Stan's implementation is an adaptive step size is used by comparing the solution obtained using the four term approximation of above as well as a five term approximation. If these approximations agree, the algorithm proceeds, otherwise a new step size is calculated. More details can be found in the user manual \citep{stan2020}, but the result is a fast, efficient procedure of high accuracy.

\subsection{Hamiltonian Monte Carlo and the No-U-Turn Sampler}
Our choice of implementation in Stan, as opposed to more traditional BUGS or JAGS \citep{Plummer03}, is pragmatic. First, the latter languages are declarative and built upon graphical models. In contrast, Stan is a fully imperative programming language. Additionally, built-in differential equation routines are included such as the Runge-Kutta numerical solver described in the previous section. This makes the software implementation of our model more natural and readable. More importantly, however, the parameters in a nonlinear compartmental model are often highly correlated, as demonstrated in below in Figure \ref{fig:cor}.

\begin{figure}[H]
    \centering
    \subfloat{
        \includegraphics[scale=.2]{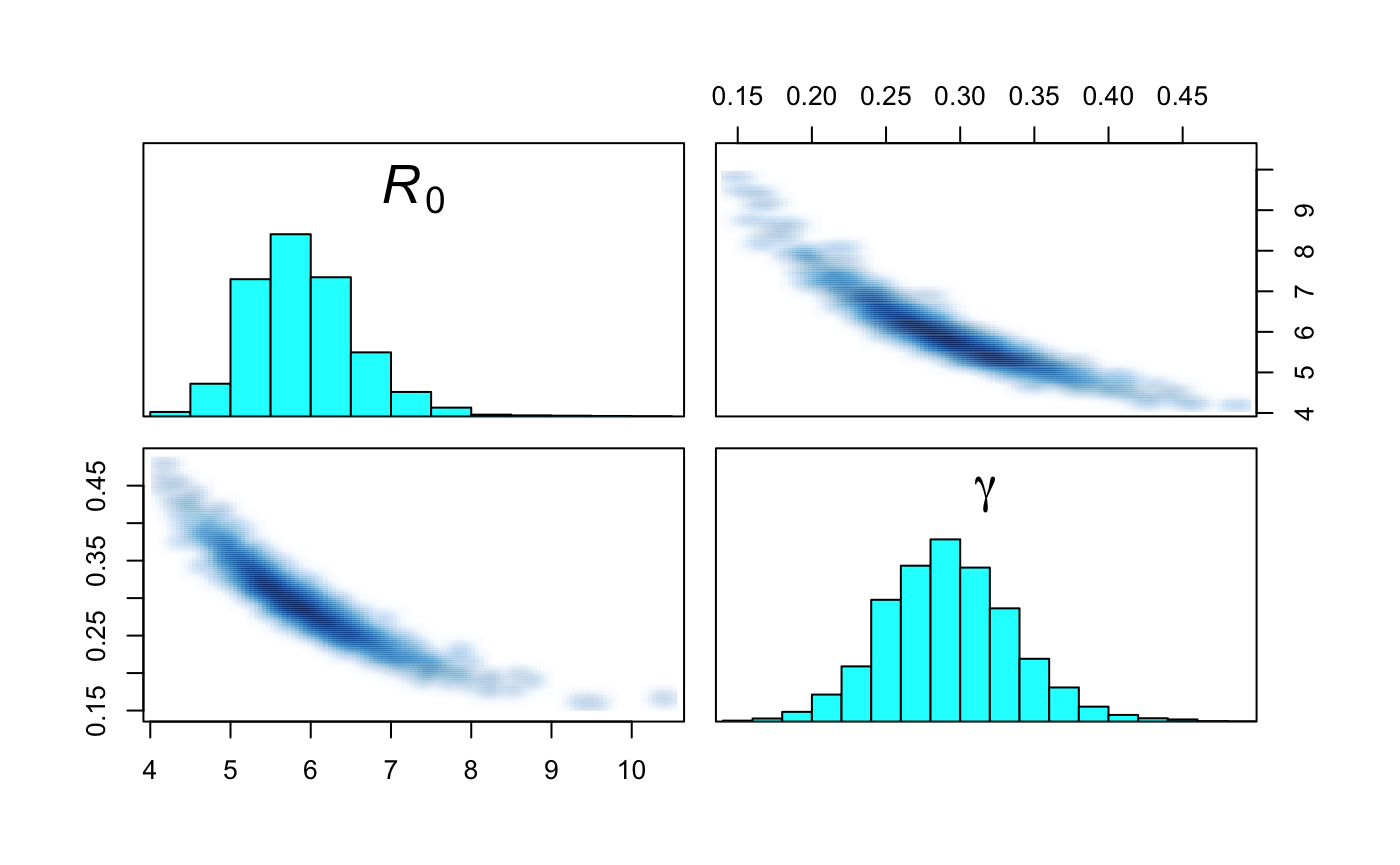}
    }
    \centering
     \caption{Posterior correlation between $\mathcal{R}_0$ and $\gamma$ due to Nonlinear Mechanistic System}
  \label{fig:cor}
\end{figure}

Hamiltonian Monte Carlo (HMC) is more equipped to sample from complex posterior distributions with high autocorrelations than standard Metropolis schemes. The most popular presentation of Hamiltonian Monte Carlo is by way of analogy with statistical mechanics. Let $\boldsymbol\theta$ be an arbitrary $d$-dimensional parameter vector. To sample efficiently from the posterior of $\boldsymbol{\theta}$ after conditioning on data, an idealized physical system is introduced to leverage the geometry of the underlying manifold on which $\boldsymbol{\theta}$ lives. We will not embark upon a comprehensive development of HMC here, referring the reader to excellent introductory expository articles by \citep{neal2012} and \citep{betancourt2018conceptual}. Instead, we provide a heuristic account of the HMC algorithm and why it works.  

We begin by recalling the more conspicuous Metropolis random walk and the concept of detailed balance. Let $p(\boldsymbol{\theta}\given \mathbf{Y})$ be the posterior distribution from which we wish to sample. As it may be difficult to directly sample from $p(\boldsymbol{\theta}\given \mathbf{Y})$, the Metropolis random-walk algorithm constructs a Markov chain with $p(\boldsymbol{\theta}\given \mathbf{Y})$ as its stationary distribution. Given an initial value $\boldsymbol{\theta}^{(0)}$, at iteration $t$ we draw a ``proposed'' value $\boldsymbol{\theta}^{\ast}$ from a symmetric distribution $q(\cdot\given \boldsymbol{\theta}^{(t-1)})$. A simple yet effective choice for many applications is $q(\cdot\given \boldsymbol{\theta}^{(t-1)}) = N(\cdot\given \boldsymbol{\theta}^{(t-1)}, \mathbf{V})$, where $\mathbf{V}$ is a fixed variance covariance matrix that helps tune the algorithm, although in general the proposal can be generated from any symmetric distribution. After generating $\boldsymbol{\theta}^{\ast}$ we simulate a coin with probability of heads $\displaystyle \min\left(1, \frac{p(\boldsymbol{\theta}^{\ast}\given \mathbf{Y})}{p(\boldsymbol{\theta}^{(t-1)}\given \mathbf{Y})}\right)$ and set $\boldsymbol{\theta}^{(t)} = \boldsymbol{\theta}^{\ast}$ if it is a head. Otherwise we set $\boldsymbol{\theta}^{(t)} = \boldsymbol{\theta}^{(t-1)}$. That $p(\boldsymbol{\theta}\given \mathbf{Y})$ is indeed the desired stationary distribution can be seen as follows. Let us assume that the current state $\boldsymbol{\theta}^{(t-1)} = \mathbf{a}$ is a draw from $p(\boldsymbol{\theta}\given \mathbf{Y})$ and consider the possibility of moving to $\boldsymbol{\theta}^{(t)}=\mathbf{b}$. This conditional probability is given by the transition probability that a value of $\mathbf{b}$ is proposed from $q(\cdot \given \mathbf{a})$ and that this value is accepted. Hence,
%$T(\mathbf{a} \to \mathbf{b}) = P(\boldsymbol{\theta}^{(t+1)}=\mathbf{b}\given \boldsymbol{\theta}^{(t)}=\mathbf{a})$
\begin{equation}\label{eqn: metrop_transition_prob}
 \begin{split}
 T(\mathbf{a} \to \mathbf{b}) = P(\boldsymbol{\theta}^{(t)}=\mathbf{b}\given \boldsymbol{\theta}^{(t-1)}=\mathbf{a}) = q(\mathbf{b}\given \mathbf{a})\min\left(1, \frac{p(\mathbf{b}\given \mathbf{Y})}{p(\mathbf{a}\given \mathbf{Y})}\right) = \min\left(q(\mathbf{b}\given \mathbf{a}), q(\mathbf{b}\given \mathbf{a})\frac{p(\mathbf{b}\given \mathbf{Y})}{p(\mathbf{a}\given \mathbf{Y})}\right)\;.
 \end{split}
\end{equation}
The form of the transition probability in  \eqref{eqn: metrop_transition_prob} implies time-reversibility (or detailed balance) in the following sense:
\begin{equation}\label{eqn: time_reversibility}
 \begin{split}
  P(\boldsymbol{\theta}^{(t-1)} = \mathbf{a}, \boldsymbol{\theta}^{(t)} = \mathbf{b}) &= P(\boldsymbol{\theta}^{(t-1)} = \mathbf{a}) T(\mathbf{a} \to \mathbf{b}) = p(\mathbf{a}\given \mathbf{Y})\min\left(q(\mathbf{b}\given \mathbf{a}), q(\mathbf{b}\given \mathbf{a})\frac{p(\mathbf{b}\given \mathbf{Y})}{p(\mathbf{a}\given \mathbf{Y})}\right) \\ 
 &= \min\left(p(\mathbf{a}\given \mathbf{Y})q(\mathbf{b}\given \mathbf{a}), q(\mathbf{b}\given \mathbf{a})p(\mathbf{b}\given \mathbf{Y})\right) = \min\left(p(\mathbf{a}\given \mathbf{Y})q(\mathbf{a}\given \mathbf{b}), q(\mathbf{a}\given \mathbf{b})p(\mathbf{b}\given \mathbf{Y})\right) \\
 &= p(\mathbf{b}\given \mathbf{Y})q(\mathbf{a}\given \mathbf{b})\min\left(\frac{p(\mathbf{a}\given \mathbf{Y})}{p(\mathbf{b}\given \mathbf{Y})}, 1\right) = P(\boldsymbol{\theta}^{(t-1)}=\mathbf{b}) T(\mathbf{b} \to \mathbf{a}) \\ 
 &= P(\boldsymbol{\theta}^{(t-1)} = \mathbf{b}, \boldsymbol{\theta}^{(t)} = \mathbf{a})\; ,
 \end{split}
\end{equation}
where we have used the symmetry $q(\mathbf{a}\given \mathbf{b}) = q(\mathbf{b}\given \mathbf{a})$ in the fourth equality in (\ref{eqn: time_reversibility}). It follows that the draw of $\boldsymbol{\theta^{(t)}}$ is also from $p(\boldsymbol{\theta}\given \mathbf{Y})$ because
\begin{equation}\label{eqn: stationarity_metropolis}
 P(\boldsymbol{\theta}^{(t)} = \mathbf{b}) = \int P(\boldsymbol{\theta}^{(t-1)} = \mathbf{a}, \boldsymbol{\theta}^{(t)} = \mathbf{b})d\mathbf{a} = \int P(\boldsymbol{\theta}^{(t-1)} = \mathbf{b}, \boldsymbol{\theta}^{(t)} = \mathbf{a})d\mathbf{a} =  P(\boldsymbol{\theta}^{(t-1)} = \mathbf{b}) = p(\mathbf{b}\given \mathbf{Y})\;.
\end{equation}

The underlying idea behind HMC is that instead of generating the proposed value from a random distribution, we use a deterministic \emph{symplectic integrator} to propose $\boldsymbol{\theta}^{\ast}$. This symplectic integrator is designed based upon Hamiltonian dynamics. Suppose that we wish to sample from $p(\boldsymbol{\theta}\given \mathbf{Y})$, where $\boldsymbol{\theta}\in \mathbb{R}^d$. We introduce an auxiliary variable $\mathbf{r}\in \mathbb{R}^d$ so that we can efficiently sample from the joint density $p(\boldsymbol{\theta}, \mathbf{r}\given \mathbf{Y})$. If $(\boldsymbol{\theta}^{(t)}, \mathbf{r}^{(t)})\sim p(\boldsymbol{\theta}, \mathbf{r}\given \mathbf{Y})$, then
\[
 P(\boldsymbol{\theta}^{(t)} = \mathbf{b}) = \int P(\boldsymbol{\theta}^{(t)} = \mathbf{b}, \mathbf{r}^{(t)}=\mathbf{u})d\mathbf{u} = \int p(\mathbf{b}, \mathbf{u}\given \mathbf{Y})d\mathbf{u} = p(\mathbf{b}\given \mathbf{Y})\;.
\]
Hence, sampling from the joint density $p(\boldsymbol{\theta}, \mathbf{r}\given \mathbf{Y})$ results in samples from $p(\boldsymbol{\theta}\given \mathbf{Y})$.

The auxiliary variable, $\mathbf{r}$, is also called the ``momentum'' in Hamilton dynamics. For our purposes, it suffices to specify that $\displaystyle p(\boldsymbol{\theta}, \mathbf{r}\given \mathbf{Y}) = p(\boldsymbol{\theta}\given \mathbf{Y})\times p(\mathbf{r})$. Therefore, $p(\mathbf{r}\given \boldsymbol{\theta}, \mathbf{Y}) = p(\mathbf{r})$ which means that $\mathbf{r}$ is independent of the data $\mathbf{Y}$ and the model parameters $\boldsymbol{\theta}$. More specifically, we assume that $p(\mathbf{r}) = N(\mathbf{r}\given \mathbf{0}, \mathbf{I}_d) \propto \exp\left(-\frac{1}{2}\mathbf{r}^{\top}\mathbf{r}\right)$. Therefore,
\begin{equation}\label{eqn: HMC_joint_posterior}
\log p(\boldsymbol{\theta},\mathbf{r}\given \mathbf{Y}) = \mbox{constant} + \log p(\boldsymbol{\theta}\given \mathbf{Y}) - \frac{1}{2}\mathbf{r}^{\top}\mathbf{r}\;.   
\end{equation}
The above density can be looked upon as a physical system subject to Hamiltonian dynamics, where $\boldsymbol{\theta}$ is a particle's position in $\mathbb{R}^d$ and $\mathbf{r}$ is the particle's momentum.

In order to sample from (\ref{eqn: HMC_joint_posterior}), a simple HMC algorithm proceeds closely on the lines of the Metropolis random walk described earlier, but replaces the random generation of a proposed value for $\boldsymbol{\theta}$ by a symplectic integrator constructed from Hamiltonian dynamics. With the current state $(\boldsymbol{\theta}^{(t-1)}, \mathbf{r}^{(t-1)})$,  we begin iteration $t$ by drawing the momentum variable $\mathbf{r}^{\ast} \sim  N(\mathbf{0}, \mathbf{I}_d)$. Setting $\mathbf{r}^{(t)}=\mathbf{r}^{\ast}$ we perform $L$ steps of a symplectic integrator (also known as ``leapfrog''), where each step comprises the following:
\begin{equation}\label{eqn: leapfrog}
 \mathbf{r}^{(t+\epsilon/2)} = \mathbf{r}^{(t)} + \frac{\epsilon}{2}\nabla_{\mathbf{\theta}}{\cal L}(\boldsymbol{\theta});\quad \boldsymbol{\theta}^{(t-1+\epsilon)} = \boldsymbol{\theta}^{(t-1)} + \epsilon\mathbf{r}^{(t+\epsilon/2)};\quad \mathbf{r}^{(t+\epsilon)} = \mathbf{r}^{(t+\epsilon/2)} + \frac{\epsilon}{2}\nabla_{\mathbf{\theta}}{\cal L}(\boldsymbol{\theta})\;, 
\end{equation}
where ${\cal L}(\boldsymbol{\theta}) = \log p(\boldsymbol{\theta}\given \mathbf{Y})$. Let $\tilde{\boldsymbol{\theta}}$ and $\tilde{\mathbf{r}}$ be the output of (\ref{eqn: leapfrog}) at the end of $L$ steps. The values of $\tilde{\boldsymbol{\theta}}$ and $\tilde{\mathbf{r}}$ are considered the ``proposed'' values at iteration $t$ and accepted as $(\boldsymbol{\theta}^{(t)}, \mathbf{r}^{(t)}) = (\tilde{\boldsymbol{\theta}}, -\tilde{\mathbf{r}})$ with acceptance probability $\displaystyle \min\left(1, \frac{p(\tilde{\boldsymbol{\theta}}\given \mathbf{Y})p(\tilde{\mathbf{r}})}{p(\boldsymbol{\theta}^{(t-1)}\given \mathbf{Y})p(\mathbf{r}^{\ast})}\right)$. This last Metropolis step together with the negation of the momentum variable in the final update ensures time-reversibility as in (\ref{eqn: time_reversibility}) and, as seen in (\ref{eqn: stationarity_metropolis}), maintains $p(\boldsymbol{\theta}\given \mathbf{Y})$ as the stationary distribution.

We provide some further intuition on the time-reversibility of the simple HMC algorithm. The key to this result is that the leapfrog iteration in (\ref{eqn: leapfrog}) preserves volumes. To be slightly more precise, let ${\cal D}$ be a small region in the $(\boldsymbol{\theta}, \mathbf{r})$ space and suppose the $L$ leapfrog steps maps ${\cal D}$ to a region $\tilde{{\cal D}}$. Then ${\cal D}$ and $\tilde{{\cal D}}$ both have the same volume. We write the transition probability from ${\cal D}$ to $\tilde{{\cal D}}$ as 
\begin{equation}\label{eqn: HMC_transition}
 T({\cal D} \to \tilde{\cal D}) = (\delta V)\min\left(1, \frac{\exp(-H(\tilde{{\cal D}}))}{\exp(-H({\cal D}))}\right) = (\delta V)\min\left(1, \exp\left(-H(\tilde{{\cal D}}) + H({\cal D})\right)\right)\;,
\end{equation}
where $\delta V$ represents the volume of ${\cal D}$ and $\tilde{{\cal D}}$, and $\displaystyle \int_{{\cal D}} p(\boldsymbol{\theta}, \mathbf{r}) d\boldsymbol{\theta}d\mathbf{r} = \exp\left(-H({\cal D})\right)$. If $(\boldsymbol{\theta}^{(t-1)}, \mathbf{r}^{\ast}$ is drawn from the joint density in (\ref{eqn: HMC_joint_posterior}), then $\displaystyle P((\boldsymbol{\theta}^{(t-1)}, \mathbf{r}^{\ast})\in {\cal D}) = \int_{{\cal D}} p(\boldsymbol{\theta}, \mathbf{r}) d\boldsymbol{\theta}d\mathbf{r} = \exp\left(-H({\cal D})\right)$. Therefore,
\begin{equation}\label{eqn: HMC_detailed_balance}
 \begin{split}
  P((\boldsymbol{\theta}^{(t-1)}, \mathbf{r}^{\ast})\in {\cal D}, (\boldsymbol{\theta}^{(t)}, \mathbf{r}^{(t)})\in \tilde{{\cal D}}) &= P((\boldsymbol{\theta}^{(t-1)}, \mathbf{r}^{\ast})\in {\cal D}) T({\cal D} \to \tilde{\cal D}) \\
  &= \exp\left(-H({\cal D})\right) (\delta V)\min\left(1, \exp\left(-H(\tilde{{\cal D}}) + H({\cal D})\right)\right) \\
  &= (\delta V)\min\left(\exp\left(-H({\cal D})\right), \exp\left(-H(\tilde{{\cal D}})\right)\right) \\
  &= P((\boldsymbol{\theta}^{(t-1)}, \mathbf{r}^{\ast})\in \tilde{{\cal D}}, (\boldsymbol{\theta}^{(t)}, \mathbf{r}^{(t)})\in {\cal D})\;,  
 \end{split}
\end{equation}
where the last equality follows from the symmetry in the expression above it.

This procedure, while maintaining the stationary distribution through time-reversibility, introduces a host of complexities. Perhaps most importantly, tuning the many parameters needed in this process is inherently difficult. This motivated the development of an automatic procedure known as the No-U-Turn-Sampler (NUTS) \citep{NUTS2011}. %During the warm-up phase of NUTS, $\mathbf{M}$ is estimated, as is the step size of the leapfrog method required to efficiently explore the phase space generated by the dynamical system. 
This %warm-up estimation is accomplished 
algorithm achieves significant efficiency over the simple HMC algorithm described above
by either explicitly avoiding a U-turn to previously explored region or terminating after a pre-defined number of exploration steps. In this way, the algorithm is guaranteed to only explore new areas of the space. This efficient exploration results in typically faster convergence and higher effective sample sizes per iteration as compared to classical MCMC.

\bibliography{frankenburg}

\end{document}